\documentstyle[12pt,epsf]{article} 

\def\lsim{\mathrel{\lower2.5pt\vbox{\lineskip=0pt\baselineskip=0pt 
           \hbox{$<$}\hbox{$\sim$}}}} 
\def\gsim{\mathrel{\lower2.5pt\vbox{\lineskip=0pt\baselineskip=0pt 
           \hbox{$>$}\hbox{$\sim$}}}} 
\def\ap{a^{\prime}}

\def\app{a^{\prime\prime}}

\def\k{\kappa}

\def\d{\delta}

\def\L{\Lambda}
\def\bL{\bar{\Lambda}}

\def\e{\epsilon}

\def\tp{\tilde{\psi}}
\def\e{\epsilon}
\def\r{\rho}
\begin{document} 
\begin{flushright}
DPNU-02-08\\ hep-th/0204113
\end{flushright}

\vspace{10mm}

\begin{center}
{\Large \bf 
 Localized gravity on de Sitter brane in five dimensions}

\vspace{20mm}
 Masato ITO 
 \footnote{E-mail address: mito@eken.phys.nagoya-u.ac.jp}
\end{center}

\begin{center}
{
\it 
{}Department of Physics, Nagoya University, Nagoya, 
JAPAN 464-8602 
}
\end{center}

\vspace{25mm}

\begin{abstract}
 We consider a single brane embedded in five dimensions with
 zero and positive cosmological constant.
 In this setup, the existence of $dS_{4}$ brane is allowed.
 We explore the gravitational fluctuations on the brane,
 it is pointed out that the usual four-dimensional gravity can be
 reproduced by a normalizable zero mode and that the continuous massive
 modes are separated by a mass gap from zero mode.
 We derive the relation among the cosmological constant in
 the brane, the five dimensional fundamental scale and Planck scale.
 Finally we show that nonzero small observed cosmological constant
 cannot be obtained in this setup.
\end{abstract} 

\newpage 
%
%
 Recent observation has indicated that the cosmological constant
 $\L_{4}$ is a sufficiently small positive value,
 $\L_{4}\sim (10^{-3}{\rm eV})^{4}$ \cite{Perlmutter:1998np}.
 Consequently it is natural to assume that we live in a de Sitter
 universe.
 However it is difficult to explain the reason for the smallness of
 $\L_{4}$, so called cosmological constant problem.
 By using the warped extra dimension scenario by
 Randall and Sundrum \cite{Randall:1999vf},
 new suggestion for the cosmological constant problem was proposed.

 Based on the assumption that physics of four dimensions should be
 explained by higher dimensional theory, the warped braneworlds with an
 infinite extra dimension are remarkable scenarios in the field of particle
 physics.
 Ref. \cite{Randall:1999vf} has led the conclusion that the gravity can
 be localized on flat $3$-brane embedded in $AdS_{5}$,
 the usual four-dimensional gravity can be reproduced at large distance
 scale.
 This is because a normalizable zero mode exists and
 the effective four-dimensional Planck scale is finite.
 Due to fine-tuning between the bulk cosmological constant and the brane
 tension, the four-dimensional cosmological constant is vanishing. 
 Furthermore, the close relation between $AdS$/CFT correspondence and 
 the Randall-Sundrum model is explored in the last couple of years
 \cite{Maldacena:1997re,Witten:1998qj,Gubser:1998bc}.
 Moreover it is expected that the warped braneworlds provide new
 scenario in the framework of phenomenological model buildings and
 cosmologies.
 We made the extension of the five-dimensional Randall-Sundrum
 model to $(5+n)$-dimensions, and analyzed the localization of gravity and 
 the small correction terms to four-dimensional Newton's law 
 \cite{Ito:2001nc,Ito:2001fd}.  
 Thus it is important to investigate the behavior of gravity in
 the warped braneworlds
 \cite{Kaloper:1999sm,Lykken:1999nb,Giddings:2000mu,Csaki:2000fc,
 Schwartz:2000ip, Gremm:2000dj,Ito:2002nw,Karch:2000ct,Miemiec:2000eq,
 Brevik:2002yj}.
 For instance, Karch and Randall had explored the behavior of the gravity 
 on $dS_{4}$ brane and $AdS_{4}$ brane embedded in $AdS_{5}$ background
 \cite{Karch:2000ct}.
 In the case of $dS_{4}$, a normalizable zero mode of gravity exists.
 In the case of $AdS_{4}$, zero mode don't exist and
 massive modes have discrete eigenvalues due to the box-like volcano 
 potential in Schr$\ddot{\rm o}$dinger equation.

 Thus an important feature of the warped braneworld is the localization of
 gravity on the brane.
 Moreover, it is interesting to study whether nonzero small
 cosmological constant $\L_4$ in brane can be obtained or not.

 In this letter, we consider the braneworlds with zero and positive
 bulk cosmological constant in contrast to Karch and Randall
 \cite{Karch:2000ct}.
 Solving the Einstein equation of this setup, it is shown that there
 only exists the solution of $dS_{4}$ brane.
 We examine the resultant volcano potential in order to study the
 fluctuations of linearized gravity on a single $dS_{4}$ brane.
 Furthermore, we investigate the value of the cosmological constant
 $\L_{4}$ in $dS_{4}$ brane.

 The action of the setup, a single $3$-brane with the brane
 tension $V$ is located at $y=0$, is given by 
 \begin{eqnarray}
  S=\int d^{4}x\;dy\sqrt{-G}
    \left(\;\frac{1}{2\k^{2}_{5}}{\cal R}
    -\L\;\right)-\int d^{4}x\;\sqrt{-g}\;V\,,\label{eqn1}
 \end{eqnarray} 
 where $\L$ is the cosmological constant in the bulk and $\k^{2}_{5}$ is
 the five dimensional gravitational constant.
 The ansatz for metric in five dimensions is given by
 \begin{eqnarray}
  ds^{2}=a^{2}(y)g_{\mu\nu}dx^{\mu}dx^{\nu}+dy^{2}\equiv
 G_{MN}dx^{M}dx^{N}\,,\label{eqn2}
 \end{eqnarray}
 where $a(y)$ is a warp factor, it is assumed that the fifth dimension $y$
 satisfies the $Z_{2}$ symmetry.
 The four-dimensional slice
 has a de Sitter metric as follows,
 \begin{eqnarray}
  g_{\mu\nu}dx^{\mu}dx^{\nu}=-dx^{2}_{0}+e^{2\sqrt{\bL}\;x_{0}}
  \left(dx^{2}_{1}+dx^{2}_{2}+dx^{2}_{3}\right)\,.\label{eqn3}
 \end{eqnarray}
 Here we define $3\bL\equiv \L_{4}/M^{2}_{\rm pl}$, where $\L_{4}$ and
 $M_{\rm pl}$ are four-dimensional cosmological constant and Planck
 scale, respectively.
 Moreover, Anti-de Sitter slices can be obtained by the following
 transformations:
 $x_{0}\rightarrow ix_{3}$, $x_{3}\rightarrow ix_{0}$, and
 $\sqrt{\bL}\rightarrow i\sqrt{\bL}$.
 Einstein equations of this setup are explicitly expressed as
 \begin{eqnarray}
  \left(\frac{\ap}{a}\right)^{2}-\frac{\bL}{a^{2}}=-\frac{\k^{2}_{5}}{6}\L
  \label{eqn4}
 \end{eqnarray}
 and
 \begin{eqnarray}
  \frac{\app}{a}
  =-\frac{\k^{2}_{5}}{6}\L-\frac{\k^{2}_{5}}{3}V\d(y)\label{eqn5}\,.
 \end{eqnarray}
 Using the above equations, we can obtain the warp factor for $\L<0$,
 $\L=0$, and $\L>0$.
 The solutions to the equations were first investigated in
 Ref. \cite{Kaloper:1999sm} which investigated the Einstein equations
 for a single brane or two branes embedded in the curved background.

 For $\L<0$, the solutions to $dS$ $(\bL>0)$ brane and $AdS$ $(\bL<0)$
 brane have been analyzed by Karch and Randall \cite{Karch:2000ct}.
 Moreover, the case of $\bL=0$ corresponds to the original
 Randall-Sundrum model \cite{Randall:1999vf}.
 For this reason, in this letter, we don't discuss the localization of
 gravity in the braneworlds with negative bulk cosmological constant.

 For $\L=0$, the solution of $AdS$ $(\bL<0)$ brane cannot exist.
 Although the solution of $\bL=0$ is allowed, the warp factor becomes
 a constant.
 Namely, the case is not interesting because of non-warped metric.
 However, the solution of $dS$ $(\bL>0)$ brane exists.
 The warp factor is given by
 \begin{eqnarray}
  a(y)=\sqrt{\bL}\left(c-|y|\right)\,,\label{eqn6}
 \end{eqnarray}
 where $c$ is a positive integration constant.
 From Eq.(\ref{eqn5}), the jump condition at $y=0$ leads to the brane
 tension
 \begin{eqnarray}
  V=\frac{6}{\k^{2}_{5}c}\,.\label{eqn7}
 \end{eqnarray}
 The form of warp factor in Eq.(\ref{eqn6}) yields the system of
 positive tension brane.

 For $\L>0$, the solution of $dS$ $(\bL>0)$ brane is allowed.
 The warp factor can be expressed as \cite{Kaloper:1999sm}
 \begin{eqnarray}
  a(y)=L\sqrt{\bL}\sin\frac{c-|y|}{L}\,,\hspace{1cm}
  L=\sqrt{\frac{6}{k^{2}_{5}\L}}\,.\label{eqn8}
 \end{eqnarray}
 From Eq.(\ref{eqn4}), note that the solutions of $\bL=0$ and $\bL<0$
 cannot be allowed.
 Furthermore, the brane tension takes the form as follows 
 \begin{eqnarray}
  V=\frac{6}{\k^{2}_{5}L}\cot\frac{c}{L}\,.\label{eqn9}
 \end{eqnarray}
 Thus the positive tension brane can be obtained by taking positive
 constant $c$.  
 From Eqs.(\ref{eqn6}) and (\ref{eqn8}), it implies that $c$ is the
 distance between the brane and the horizon.
 Moreover, the normalization condition of $a(0)=1$ leads to the
 relation between $\bL$ and $c$.
 Consequently, we have
 $\bL=c^{-2}$ for $\L=0$, and $\bL=L^{-2}\sin^{-2}c/L$ for $\L>0$.
 From Eqs.(\ref{eqn7}) and (\ref{eqn9}), we have
 $\bL=(\k^{2}_{5}V/6)^{2}$ for $\L=0$ and
 $\bL=(\k^{2}_{5}V/6)^{2}+L^{-2}$ for $\L>0$.
 Taking the limit of large $L$ (small $\L$), the warp factor of
 Eq.(\ref{eqn8}) approaches the one of Eq.(\ref{eqn6}).
 For the limit of $c/L\ll 1$, Eq.(\ref{eqn9}) is consistent with
 Eq.(\ref{eqn7}).  

 As mentioned above, in setup considered here,
 it turns out that a $dS_{4}$ brane can be allowed in five dimensions with
 zero and positive cosmological constant.
 Below, in order to study the behavior of gravity we examine the
 potential in wave equation for gravitational fluctuation.

\begin{figure}
      \epsfxsize=7.5cm
\centerline{\epsfbox{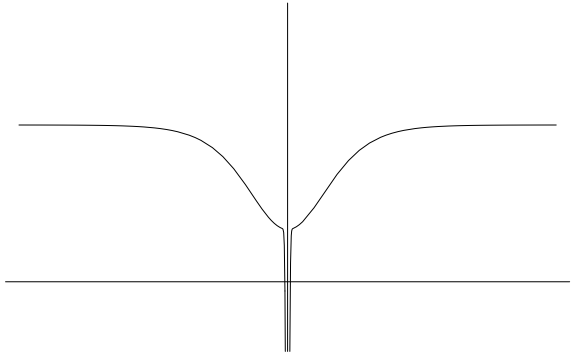}}
\caption{Volcano potential in $dS$ brane for $\L>0$}
\label{fig1} 
\end{figure}

 Taking $g_{\mu\nu}+a^{-2}(y)h_{\mu\nu}(x,y)$ as the gravitational
 fluctuation around the background metric, we can obtain the
 wave equation for gravity.
 By imposing transverse-traceless gauge conditions for these
 fluctuations, a separation such as $h_{\mu\nu}(x,y)=h_{\mu\nu}(x)\psi(y)$ 
 yields the wave function for gravity as follows
 \begin{eqnarray}
  \frac{d^{2}}{dy^{2}}\psi
  +\left[-2\frac{\app}{a}-2\left(\frac{\ap}{a}\right)^{2}
  +m^{2}a^{-2}\right]\psi=0\,.\label{eqn10}
 \end{eqnarray}
 Here we used wave equation of the four-dimensional gravity in curved
 background,
 $(\Box +2\L_{4}/M^{2}_{\rm pl})h_{\mu\nu}=m^{2}h_{\mu\nu}$,
 where $\Box$ is the four-dimensional covariant d'Alembertian and 
 $m^{2}$ corresponds to the mass of four-dimensional gravity.
 Transforming to conformally flat coordinate $z=\int dy\;a^{-1}(y)$
 and performing replacement of $\psi(y)=a^{1/2}(y)\tp (z)$,
 we have the familiar Schr$\ddot{\rm o}$dinger equation.
 For the cases of $\L=0$ and $\L>0$, we show that the relation 
 between $y$-coordinate and $z$-coordinate and the resultant volcano
 potential in the Schr$\ddot{\rm o}$dinger equation.
 The conformally $z$ coordinate is given by  
 \begin{eqnarray}
 \L=0&:&z(y)=sgn(y)\frac{1}{\sqrt{\bL}}\label{eqn11}
 \left\{\log\left(\frac{1}{c-|y|}\right)-z_{0}\sqrt{\bL}\right\}\\
 \L>0&:&z(y)=sgn(y)\frac{1}{\sqrt{\bL}}
 \left\{{\rm arc}\cosh\left(\frac{1}{\sin\frac{c-|y|}{L}}\right)
 -z_{0}\sqrt{\bL}\right\}\label{eqn12}\,,
 \end{eqnarray}
 where the constant $z_{0}$ is given by
 \begin{eqnarray}
 \L=0&:&z_{0}=\frac{1}{\sqrt{\bL}}\log\frac{1}{c}\label{eqn13}\\
 \L>0&:&z_{0}=\frac{1}{\sqrt{\bL}}{\rm arc}\cosh
 \left(\frac{1}{\sin\frac{c}{L}}\right)\,.\label{eqn14}
 \end{eqnarray}
 Thus the $z$ coordinate runs from $-\infty$ to $+\infty$.
 The warp factors can be rewritten in terms of $z$, accordingly,
 $a(z)=e^{-\sqrt{\bL}|z|}$ for $\L=0$ and
 $a(z)=L\sqrt{\bL}/\cosh(\sqrt{\bL}(z_{0}+|z|))$ for $\L>0$.
 Using Eqs.(\ref{eqn6}) and (\ref{eqn8}), Eq.(\ref{eqn10}) is expressed 
 in terms of $z$
 \begin{eqnarray}
  \left[\;
  -\frac{d^{2}}{dz^{2}}+V(z)\;\right]\tp=m^{2}\tp\,,\label{eqn15}
 \end{eqnarray} 
 where the volcano potential for $\L=0$ and $\L>0$ is given by 
 \begin{eqnarray}
 \L=0&:& V(z)=\frac{9}{4}\bL-3\sqrt{\bL}\d(z)\label{eqn16}\\
 \L>0&:& V(z)=\frac{9}{4}\bL-\frac{15}{4}
 \frac{\bL}{\cosh^{2}\sqrt{\bL}\left(z_{0}+|z|\right)}
 -3\sqrt{\bL}\tanh z_{0}\sqrt{\bL}\;\d(z)\label{eqn17}\,.
 \end{eqnarray}
 In the case of $\L=0$ the volcano potential is a positive constant.
 In the case of $\L>0$ as shown in Fig.\ref{fig1},
 the potential approaches a non-zero constant $9\bL/4$ for
 $|z|\rightarrow\infty$.
 Note that there is no blow-up at the horizon in
 contrast to setup of $AdS_{4}$ brane embedded in $AdS_{5}$ analyzed
 by Karch and Randall \cite{Karch:2000ct}.
 Since the Hamiltonian in Eq.(\ref{eqn15}) is a positive definite value,
 the eigenvalue $m^{2}\geq 0$ \cite{Csaki:2000fc}.

 For $\L=0$, solving the Schr$\ddot{\rm o}$dinger equation imposed
 by the jump condition at $z=0$, the zero mode wave function is given by 
 $\tp_{0}(z)=(3\sqrt{\bL}/2)^{1/2}e^{-\frac{3}{2}\sqrt{\bL}|z|}$.
 Since there is a normalizable zero mode wave function, 
 the localization of gravity occurs on the brane.
 Namely, the effective four-dimensional Planck scale is  finite value.
 The massive mode wave functions for $0<m^{2}<9\bL/4$ are given by
 $\tp_{m}\sim {\rm exp}[{-\sqrt{9\bL/4-m^{2}}\;|z|}]$,
 however, these modes do not exist there.
 This is because these modes aren't consistent with jump condition.
 For $m^{2}\geq 9\bL/4$, the wave functions become plane waves
 with continuous modes.
 Thus a zero mode mode is separated by a mass gap $9\bL/4$ from
 the continuous massive modes.
 The gravitational potential between two unit masses
 via zero mode is given by 
 $V(r)\sim \k^{2}_{5}|\tp_{0}(0)|^{2}/r= 3\k^{2}_{5}\sqrt{\bL}/2r$.
 The contributions of massive modes are described in summary.
 Therefore we can obtain the effective four-dimensional Planck scale
 $M^{-2}_{\rm pl}= 3\k^{2}_{5}\sqrt{\bL}/2$. 
 Consequently, the four-dimensional cosmological constant
 $\L_{4}$ is given by
 \begin{eqnarray}
 \L_{4}=\frac{4}{9}M^{4}_{\rm pl}
 \left(\frac{M}{M_{\rm pl}}\right)^{6}\,.
 \label{eqn18}
 \end{eqnarray}
 Here we used $\bL\sim \L_{4}/M^{2}_{\rm pl}$ and 
 $\k^{2}_{5}\equiv M^{-3}$, where $M$ is the fundamental scale of
 five dimensions.
 Note that $\L_{4}$ is in proportion to sixth powers of ratio
 $M/M_{\rm pl}$,
 Eq.(\ref{eqn18}) is consistent with $\L_{4}$ in warped
 compactification of Type IIB string theory \cite{Berglund:2002kr}.
 The observed value of $\L_{4}$ leads to
 the ratio $M/M_{\rm pl}\sim 10^{-20}$, namely, we obtain
 $M={\cal O}(100\;{\rm MeV})$.
 Thus this case is phenomenologically ruled out.
 Setting $M={\cal O}(\rm TeV)$, the cosmological constant in brane cannot
 be consistent with the observed value.

 For $\L>0$, the general solution of the Schr$\ddot{\rm o}$dinger equation 
 with potential of Eq.(\ref{eqn17}) is expressed as
 \begin{eqnarray}
 &&\tp_{m}(z)=A\left[\cosh \sqrt{\bL}(z_{0}+|z|)\right]^{-\r}
 {}_{2}F_{1}\left(\r-\frac{3}{2},\r+\frac{5}{2};1+\r ;
 \frac{1-\tanh \sqrt{\bL}(z_{0}+|z|)}{2}\right)\nonumber\\
 &&+B\left[\cosh \sqrt{\bL}(z_{0}+|z|)\right]^{\r}
 {}_{2}F_{1}\left(-\r-\frac{3}{2},-\r+\frac{5}{2};1-\r ;
 \frac{1-\tanh \sqrt{\bL}(z_{0}+|z|)}{2}\right)\label{eqn19}
 \end{eqnarray}
 where $\r=\sqrt{9/4-m^{2}/\bL}$ and $A,B$ are the normalization factors.
 Taking account into the divergence at $|z|\rightarrow\infty$,
 both $B=0$ and $3/2-\r=n$ can be required by the finiteness of
 hypergeometric function, where $n\in{\bf Z}^{+}_{0}$.
 Namely, the cases of $n=0,1$ are allowed.
 For $n=0$ $(\r=3/2, m^{2}=0)$,
 this corresponds to a zero mode of gravity and the corresponding
 wave function is given by $\tp_{0}(z)\sim
 \cosh^{-3/2}\sqrt{\bL}(z_{0}+|z|)$, which
 is consistent with the jump condition at $z=0$.
 For $n=1$ $(\r=1/2, m^{2}=2\bL)$, the massive modes take the following
 form, 
 $\tp_{m}\sim \sinh\sqrt{\bL}(z_{0}+|z|)\cosh^{-3/2}\sqrt{\bL}(z_{0}+|z|)$,
 however, the wave function isn't consistent with the jump condition. 
 Namely, in the region of $0\leq m^{2}<9\bL/4$, there is a normalizable
 zero mode.
 For $m^{2}>9\bL/4$, since $\r$ is imaginary number, the solutions of
 Eq.(\ref{eqn19}) are plane waves with continuous modes at large $z$.
 Also in this case there exists a mass gap between the zero mode and
 massive modes, the onset of massive mode starts from
 $\sqrt{9\bL/4}$.
 The behavior of Newton potential via massive modes is described in
 summary.
 At $r\gg r_{0}=2/3\sqrt{\bL}$, a zero mode generates the usual
 Newton's law.
 Estimating the normalization factor of zero mode wave function,
 we can obtain the cosmological constant $\L_{4}$ in $dS$ brane as follows,
 \begin{eqnarray}
 \L_{4}= M^{4}_{\rm pl}\left(\frac{M}{M_{\rm pl}}\right)^{6}
 (1+\e^{2})^{3}\left(\frac{\pi}{2}-\frac{\e}{1+\e^{2}}
 -{\rm arc}\tan \e\right)^{2}
 \,,\hspace{0.5cm}
 \e\equiv \frac{LV}{6M^{3}}\,.\label{eqn20}
 \end{eqnarray} 
 Note that the value of $\L_{4}$ is determined by the bulk
 curvature $L$ when $M$ and $V$ are fixed.
 For instance, the limit of $\e\rightarrow\infty$ leads to
 \begin{eqnarray}
 \L_{4}= \frac{4}{9}M^{4}_{\rm pl}
 \left(\frac{M}{M_{\rm pl}}\right)^{6}\,,
 \label{eqn21}
 \end{eqnarray}
 where this situation corresponds to the case of $\L=0$.
 Moreover the case of $\e=0\;(\L\rightarrow\infty)$ leads to
 \begin{eqnarray}
 \L_{4}= \frac{\pi^{2}}{4}M^{4}_{\rm pl}
 \left(\frac{M}{M_{\rm pl}}\right)^{6}\,.
 \label{eqn22}
 \end{eqnarray}
 Thus, for arbitrary value of the cosmological constant in the bulk,
 we obtain approximately
 $\L_{4}\sim {\cal O}(1)\times M^{4}_{\rm pl}(M/M_{\rm pl})^{6}$.
 Similar to the case of $\L=0$, we must set $M={\cal O}(100\;{\rm MeV})$
 in order to obtain the observed value of $\L_{4}$.
  
 In summary, we considered a single positive tension $3$-brane embedded in
 five-dimensional world with zero $(\L=0)$ and positive $(\L>0)$
 cosmological constant.
 Solving Einstein equations of the setup, $dS$ brane ($\bL>0$) is allowed,
 and the cases of flat brane ($\bL=0$) and $AdS$ brane ($\bL<0$) are
 ruled out.
 Furthermore, we investigated whether nonzero small cosmological
 constant in brane can be obtained or not.

 In the case of $dS$ brane in $\L=0$, the resultant volcano potential
 is a positive nonzero constant, and positive tension brane gives rise
 to a delta function of attractive force.
 Furthermore, there is a mass gap between zero mode and continuous
 massive mode.
 We derived the relation between the cosmological constant in the
 brane and the five-dimensional fundamental scale.
 In order to obtain nonzero small observed cosmological constant, 
 it turns out that the five-dimensional fundamental scale must be the
 order of a few MeV scale.

 In the case of $dS$ brane in $\L>0$, the resultant volcano potential
 approaches a positive non-zero constant at large conformally coordinate
 $z$.
 In similar to the above case, there exists a mass gap between zero mode
 and massive modes.
 It is shown that the cosmological constant in brane depends on the
 value of bulk curvature.
 Similar to the case of $\L=0$, nonzero small observed cosmological
 constant is tuned by the five dimensional scale of a few MeV scale.

 We must describe some comments with respect to the behaviors of Newton
 potential via the massive modes.
 For $\L=0$ and $\L>0$, taking into account of the massive modes,
 the Newton potential on the $dS$ brane were investigated in 
 Ref.\cite{Kehagias:2002qk}.
 The continuous massive modes separated by the mass gap from zero mode
 give the correction terms to the Newton potential.
 At the distance larger than the inverse of lowest massive mode and
 smaller than Hubble scale $H=\sqrt{\bL}$, the five dimensional gravity
 appears on the brane due to dominance of correction terms.

 Thus the localization of gravity on the brane occurs in the setup
 considered here, however, nonzero small observed cosmological
 constant cannot be obtained.
 The setup of $dS$ brane embedded in $\L=0$ and $\L>0$ can be
 ruled out from the viewpoint of astrophysics measurements.
 Here we give some comments.
 Although de Sitter bulk may be unusual in the cosmological context, 
 there are a number of investigations of $dS$/CFT correspondence
 inspired by $AdS$/CFT.
 There are many works exploring fully satisfactory $dS$
 solution of string theory, it is expected that $dS$ may provide
 several solutions to the cosmological constant problem or quantum gravity.
%
%
 
\end{document}